# A Survey of Memristive Threshold Logic Circuits

Akshay Kumar Maan, Deepthi Anirudhan Jayadevi, and Alex Pappachen James, *Senior Member, IEEE*

*Abstract*—In this paper, we review the different memristive threshold logic (MTL) circuits that are inspired from the synaptic action of flow of neurotransmitters in the biological brain. Brain-like generalisation ability and area minimisation of these threshold logic circuits aim towards crossing the Moore's law boundaries at device, circuits and systems levels. Fast switching memory, signal processing, control systems, programmable logic, image processing, reconfigurable computing, and pattern recognition are identified as some of the potential applications of MTL systems. The physical realization of nanoscale devices with memristive behaviour from materials like $TiO_2$, ferroelectrics, silicon, and polymers has accelerated research effort in these application areas inspiring the scientific community to pursue design of high speed, low cost, low power and high density neuromorphic architectures.

## I. INTRODUCTION

SEARCH for scalable hardware architectures for emulating neuron circuits in the biological brain is one of major research areas to realising silicon based artificial intelligence devices. The implementation of such circuits in standard devices such as CMOS has limitations of design and scalability. Memristive devices provide an interesting alternative not only by way of their high packing density, but also the way they can enable a profoundly different approach to large-scale computing inspired by the principle of firing of neurons in the biological brain. In biological brain a neuron fires only if the total weight of the synapses that receive impulses in a short period (called the period of latent summation) exceeds a threshold. A threshold logic implemented in memristive devices thus offers (a) a synaptic action in that the weight (memristance of the device) can be incrementally modified by controlling charge or flux through it, and (b) a thresholding system that governs the firing of the output.

Memristor, for example, is one such device [1]–[6] in the 'more-than-Moore (MtM)' era of device integration [7]–[11], that has a high packing density with features like low operational voltage, non-volatility, and high switching speed. Memristor (a portmanteau of MEMory ResISTOR) is currently being described as the nanoscale device capable of emulating synaptic behaviour in the brain [12]–[14] because it can 'remember' the charge flown through it by changing its resistance. A wide range of device models and applications have sprung up in the recent years since Hewlett-Packard (HP) Labs in 2008 [15], [16] described the behaviour of physical $TiO_2$-$TiO_{2-x}$ devices by a memristor model.

A.P. James is an Associate Professor in Electrical and Electronic Engineering Nazarbayev University. Corresponding author email: apj@ieee.org; www.biomicrosystems.info
A.J. Deepthi is a research staff member with Enview R&D labs.
A.K. Maan is a research staff at Griffith University, Australia

*Threshold logic* (TL) itself was first introduced by Warren McCulloch and Walter Pitts in 1943 [17] in the early model of an artificial neuron based on the basic property of firing of the biological neuron. The logic computed the sign of the weighted sum of its inputs. Modelling a neuron as a threshold logic gate (TLG) that fires when input reaches a threshold has been the basis of the research on neural networks (NNs) and their hardware implementation in standard CMOS logic [18]. However, to continue geometrical scaling of semiconductor components as per Moore's Law (and beyond), semiconductor industry needs to investigate a future beyond CMOS, and designs beyond the fetch-decode-execute paradigm of von Neumann architecture. Memristive devices lend themselves nicely to this "neuromorphic" computing paradigm making use of very-large-scale integration (VLSI) systems to mimic neuro-biological architectures present in the nervous system that need no initialising software, run on negligibly low power, and are able to perform many cognitive tasks rapidly, effortlessly and in real-time. Memristive threshold logic is thus a step in the direction of building neuromorphic architectures while overcoming the design and scalability issues presented by CMOS NNs.

In the following section, we give a brief background of memristive systems and threshold logic. Section III provides a review of various implementations of Memristor Threshold Logic (MTL) detailing the different architectures employed to achieve thresholding behaviour with memristive circuits and their applications, followed by discussion and open problems in section IV.

## II. BACKGROUND

### A. Memristive Devices and Systems

*Memristors* are two-terminal circuit elements exhibiting passivity property characterized by a relationship between the charge $q(t)$ and the flux-linkage $\varphi(t)$. The memristance (M) of memristor has been defined by the relation $M = d\varphi_m/dq$ and is expressed in *Wb/C*, or *ohm*. The characteristic behaviour of a memristor was proposed by L. O. Chua in 1971 [19]. He developed a circuit model built with at least 15 transistors and other passive elements [16] to emulate the behaviour of a single memristor. The memristor model satisfied the passivity criterion and was characterized by a monotonically increasing $\varphi - q$ curve [20], [21]. Fig. 1(a) depicts the relationship between the four fundamental circuit elements (Resistor (R), Capacitor (C), Inductor (L) and Memristor (M)) and the typical voltage-current behaviour of a $TiO_2$-$TiO_{2-x}$ memristive device. In 1976, L. O. Chua et al. [22] defined any nonlinear dynamical systems with memristors as *Memristive systems* . The memristive system was found to possess memory and behaved like resistive devices endowed with a variety of dynamic characteristics. The memristive systems were incapable of energy discharge. Yet, they were found to exhibit small-signal inductive or capacitive effects without introducing phase shift between the input and output waveforms.

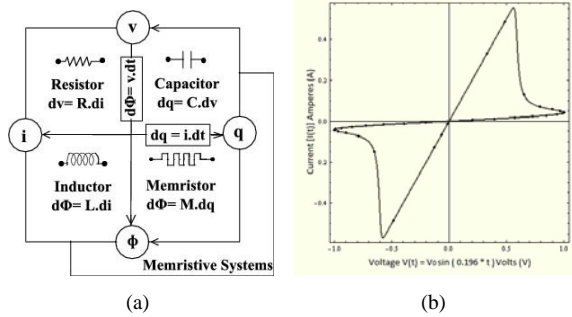

(a) (b)

Fig. 1. (a)The four fundamental two-terminal circuit elements: Resistor (R), Capacitor (C), Inductor (I) and Memristor (M). R, C, L and M can be functions of the independent variable in their defining equations, yielding nonlinear elements. For example, a charge-controlled memristor is defined by a single-valued function M(q).(b)Voltage-Current behaviour in $TiO_2 - TiO_{2-x}$ memristor model [16]

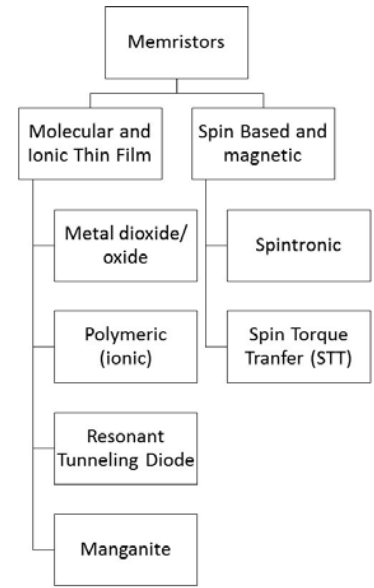

Fig. 2. A taxonomy of memrisors based on memristive behaviour observed in fabrication materials [60]

A physical solid-state device using nanoscale *Titanium dioxide* films exhibiting the memristive properties was invented in 2008 by a team from HP Labs [15]. The memristive device was found to be equipped with an ability to function like synapses in a biological brain [12]. The research team proposed a crossbar architecture which was a fully connected mesh of perpendicular platinum wires with memristive switches made of $TiO_2$-$TiO_{2-x}$ to connect any two crossing wires. These nanoscale switches were found to exhibit Lissajoux voltage-current behaviour shown in Fig. 2 (b). Till then, similar dynamics were only common in relatively larger devices [23], [24]. The memristive dynamics mapped by the pinched-hysteresis loop [25] explains switching behaviour [26] of the device beginning with a high resistance. As the applied voltage increases, the charge flow inside the device increases slowly at first owing to the drop in resistance value. This behaviour is followed by a rapid increase in the device current up to the maximum increase in applied voltage. When the voltage was decreased, the current decreased more slowly resulting in an on-switching loop. The off-switching loop was observed when the voltage turned negative leading to the increase in resistance of the device. It was observed that the resistance of the film as a whole was dependent on how much charge had been passed through it in a particular direction. In addition, the resistance value of the film was found to be reversible on changing the direction of current [19], [27], [28]. The HP device was considered as a nanionic device owing to its property of displaying fast ion conduction at nanoscale.

In 2008, D. Strukov et al. [16] presented analytical results which showed that the memristance arises naturally in nanoscale systems in which solid-state electronics and ionic transport are coupled under an external bias voltage. This paved way to set the foundation for understanding a wide range of hysteretic current-voltage behaviour observed in many nanoscale electronic devices that involved the motion of charged atomic or molecular species; in particular certain titanium dioxide cross-point switches [16], [19]–[21], [29]–[55]. In 2008, G. Chen [56] recounted the impact of invention of memristive devices in technology as recognition to the promising discovery of memristors by L.O. Chua. The modeling aspect of engineering that involved memristors and memristive systems were discussed in-depth by I. C. Goknar in 2008. In 2010, the HP Labs introduced practical memristors of size 3nm by 3nm found operable at a switching time of 1 ns (∼ 1 GHz) [57]. Memristors have thus paved way for further miniaturization of integrated electronic circuits [58], promising a future in technology beyond Moore's Law [59].

The memristive systems based on device properties can be broadly grouped into those based on molecular and ionic thin films, and into those based on spin and magnetic effects. An overall taxonomy of the memristors based on their memristive material properties [60], [61] has been presented in Fig. 2. The Ionic Thin Film and Molecular memristors mostly rely on different material properties of the thin film atomic lattices that display hysteresis below the application of charge. These memristors can be classified into four distinct groups viz. Metal Dioxide memristors, Ionic or Polymeric memristors, Resonant Tunneling Diode memristors and Manganite memristors. The metal dioxide memristors, titanium oxide in particular, are broadly explored for designing and modeling. The ionic or polymeric memristors utilize dynamic doping of inorganic die-electric type or polymer materials. In this type of memristors, the ionic charge carriers move all over the solid state structure. The resonant tunneling diode memristors use specially doped quantum well diodes of the space layers between the sources and drain regions. The manganite memristors use a substrate of bilayer oxide films based on manganite as opposed to titanium dioxide memristors.

The magnetic and spin-based memristors are opposite to ionic nanostructure and molecule based systems. This category of memristive devices rely solely on the degree of electronic spin and its polarization. This memristor type can be categorized further into two types viz. Spintronic memristors and Spin Torque Transfer (STT) memristors. In spintronic memristors, the route of spin of electrons changes the magnetization state of the device which consequently changes its resistance. In Spin Torque Transfer (STT) memristors, the comparative magnetization position of the two electrodes affect the magnetic state of a tunnel junction which in turn changes its resistance. Since HP's announcement, interest in memristive electronics and their applications has grown rapidly with several research groups demonstrating memristive behavior in different devices and systems. Fig. 3 shows the taxonomy of memristor applications proposed by Pinaki Mazumder [62] in 2012. The devices reported in literature all have different underlying physics governing their memristive behaviour.

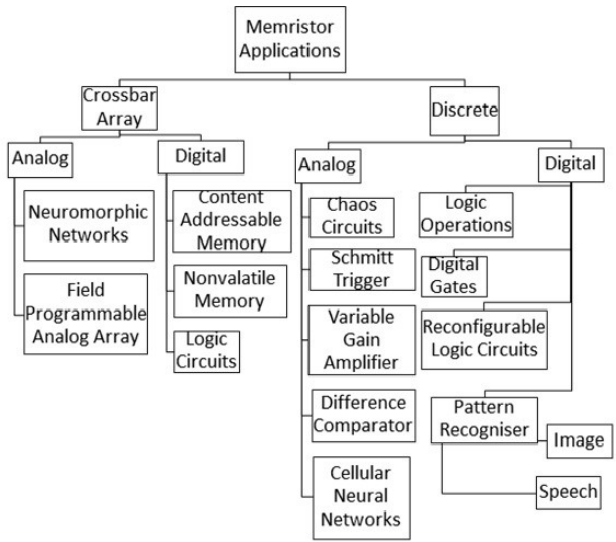

Fig. 3. A modified taxonomy of memristior circuit applications from that reported in [62] to incorporating emerging applications such as in imaging and speech.

*B. Threshold Logic*

The first simplified mathematical model of the biological neuron was introduced by McCulloch and Pitts [17] in 1943 in the form of the threshold logic gate (TLG). It computed the sign of the weighted sum of inputs:

$$f(x_1, x_2, ...x_n) = sgn(w_1x_1 + w_2x_2 + ... + w_nx_n)$$
$$= sgn(\Sigma_{i=1}^{n} w_ix_i - T) \quad (1)$$

where $w_i$ are the synaptic weights associated to inputs $x_i$ and $T$ the threshold that the gate needs to meet to fire, and $n$ is the fan-in of the TLG [63]. Since then a large number of hardware implementations have been reported in literature, a comprehensice survey of which can be found in [63]. The implementations range from early electromechanical (tubes, motors and clutches) "neurocomputer" in 1951 [64], to the potentiometer *perceptron* in 1957 [65], to the electrically adjustable *memistor* based adaptive linear element (ADALINE)

[46], and to the various very-large-scale-integration (VLSI) implementations in last two decades. These implementations are too many to be included in this survey, but they can be broadly divided into CMOS, capacitive, output-wired inverters, floating-gate and psuedo-nMOS solutions, each catering to a specific need or one or more performance parameters such as power dissipation, noise margins, sensitivity to process variations and fan-in.

The survey in [63] concludes with an interesting observation. The authors provided a list of potential applications of threshold logic apart from hardware neurons, such as " ... *general microprocessors, DSPs, and cores where addition, multiplication, and multiply-accumulate, ...encryption/decryption, ...convolution/deconvolution, ...and compression/decompression [66].*" The fundamental reason, the authors point out, is that TLGs need full custom design and have been in direct competition with Boolean gates and a lot of research effort has been spent on improving boolean logic gates since the 1970s. In this survey we will identify approaches and techniques on how to use threshold logic based on memristive systems to build different logic functionalities, which will we hope direct some of the future research effort in this direction. The authors in [63] offer another insight: "*Lastly, because nano (and reconfigurable) computing will probably get center-stage positions in the (near) future, TL will surely benefit from that.*" This observation is another motivation of the current survey because memristive threshold logic, as we shall see, is capable not only of emulating synaptic action in hardware, but it does so in a highly dense, low power and scalable architecture.

III. THRESHOLD LOGIC GATES AND APPLICATION CIRCUITS USING MEMRISTORS

The resistive switching nature exhibited by memristors [67]–[69] has been utilized in realizing brain-inspired threshold logic computing circuits. Here we present a review of the different implementations of Memristor Threshold Logic (MTL) which realizes specific function pertaining to the application under consideration.
An early implementation of threshold logic using memristors was proposed by Rajendran J. et al. [70] in 2009. The proposed Programmable Threshold Logic Array (PTLA) using memristors exhibited multiple levels of resistance to provide weighted inputs to each threshold gate. The distinguishing feature of PTLA was the combination of Negative Differential Resistance (NDR) based molecular switch and a multi-level resistance memristor leading to the programmable threshold gate. The design was extended to the implementation of an image classifier which classifies a 3x5 image into a rectangle or a triangle using three 5-input TLG in the first level and a Goto pair based majority voter as shown in Fig. 4. A Goto pair or 'twin' [71] is a series connection of two tunnel negative resistance diodes that has been used as a majority voting circuit. When the operating voltages are just enough to bring the pair to either of the stable states (0 or 1), a majority circuit can be built as in [71]

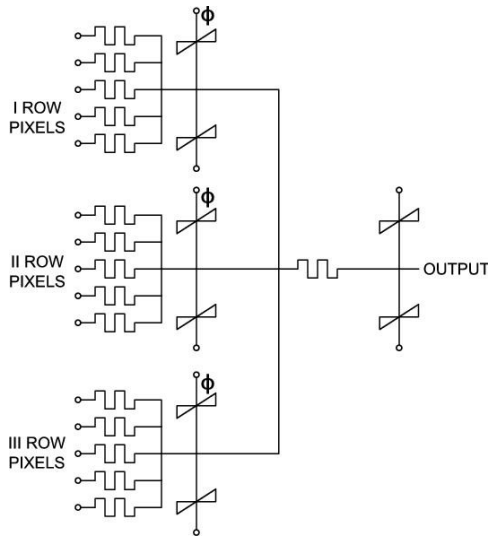

Fig. 4. TLG Implemented Using Memristors and Goto pair [70]

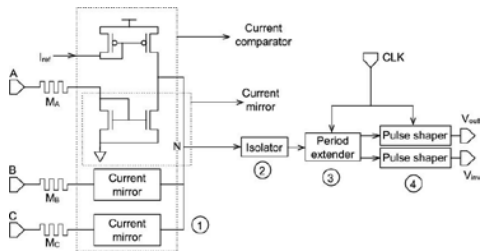

Fig. 5. Circuit diagram of a 3-input threshold gate using memristors as weights. (1) is a current mirror to prevent reverse flow of current, (2) is an isolator to prevent loading, (3) is a period extender to retain input pulse period, and (4) is a pulse shaper to retain memristance. [72]

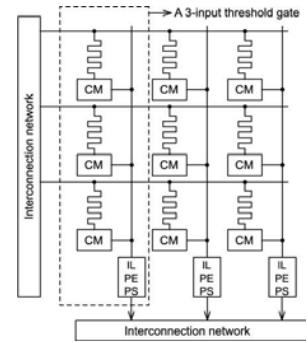

Fig. 6. An island of threshold gates [72]. Each column corresponds to a single threshold gate and the number of rows determine the fan-in of the threshold gate. A single threshold gate had been shown within a dotted box. IL refers to isolator, CM refers to current mirror, PE refers to period extender and PS refers to pulse shaper.

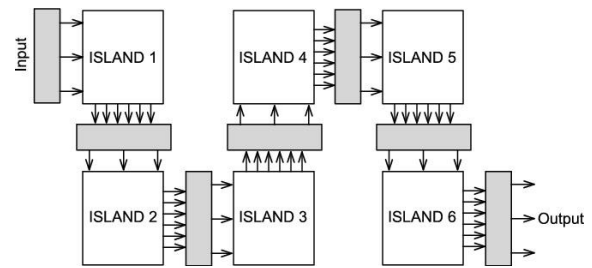

Fig. 7. A cascaded architecture using two rows of islands [72]. The unshaded box corresponds to an island of threshold gates. The lightly shaded box corresponds to an interconnection network.

to output the majority value among the inputs.

In 2010, Rajendran J. et al. [72] implemented different Boolean functions by programming the weights at the input of gates in the proposed threshold gate-array architecture. The key idea was to use the memristors as weights of the inputs to a threshold gate as shown in Fig. 5. Additional circuits were employed to avoid the problems arising from the straightforward implementation of the memristor based threshold gate, such as reverse current from output to input of the circuit, loading of the current comparator by the next stage, partial restoration of the input and temporal change in memristance. To avoid these problems, the authors added circuitry, namely: 1) a 2-transistor current mirror on each input line, 2) a 2-transistor isolation circuit, 3) a period extender circuit to restore the negative pulse immediately following the positive pulse to represent the logic 1, 4) a six transistor pulse shaper circuit to generate a positive pulse followed by a negative pulse to represent a logic 1. In general, a N-input threshold gate required $(2 \times n) + 18$ transistors. The proposed 3x3 crossbar based island architecture shown in Fig. 6 consisted of three 3-input threshold gates. A cascaded architecture formed using these threshold gate islands connected to each other through an interconnection network shown in Fig. 7. prevented signal degradation in successive stages with the help of the period extender and pulse shaper which provided signal and memristance restoration. The programming circuitry consisted of a pulse generator to program the memristor. The memristance property utilized in the architecture was able to reduce the power consumption and effective area footprint to approximately 75% in comparison with the CMOS based LUTs (Look-up Tables). The delay penalty of the programmable threshold gates was found as almost 12 times the delay of the 4-input LUTs. The memristors were utilized as weights in the realization of low-power Field Programmable Gate Arrays (FPGAs) using threshold logic in 2012 by Rajendran J. et al. [73]. The proposed Memristive Threshold Logic (MTL) gate shown in Fig. 8 utilized the multiresistance property of memristors to implement the Boolean functions, which are the subsets of threshold functions. The programmable threshold gates consume less power and area when compared to their implementations using CMOS, LUTs, and CTL gates. The energy performance and area overhead of threshold logic implementation were evaluated using the CAD tools from Cadence and Berkeley SIS logic synthesis tool. In addition, the essential countermeasures to combat the issues of using memristors in logic circuits in presence of memristance drift like memristor refresh were proposed.

Another circuit design capable of realizing four different logic operations by changing the resistance of the memristive devices was proposed by T. Tran et al. in 2012 [74]. The design exploration of reconfigurable Threshold Logic Gates (TLG) implemented using Silver-chalcogenide memristive

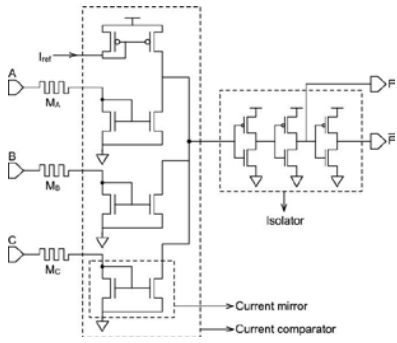

Fig. 8. A 3-input MTL gate with memristors as weights and Iref as the threshold [73]

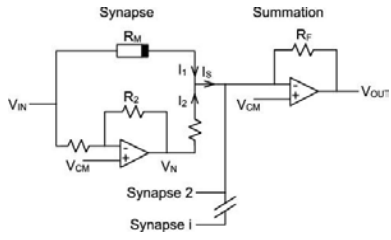

Fig. 9. Circuit implementation with memristive devices with circuitry for negative weights [74]

devices [75] combined with CMOS circuits was presented in the work. The proposed re-programmable TLG shown in Fig. 9 was realized in discrete hardware using a summing op-amp circuit with memristive devices implementing the weights ($w_i$). A feedback-based adaptive programming circuit shown in Fig. 10 was developed to program the individual memristive devices to predetermined resistance values to create each logic operation. According to the analysis carried out in Matlab-Simulink/Cadence, the TLGs with Ag-Ch memristive devices had a fan-in of >10, switched at >1 GHz speed and dissipated lower static power than a corresponding CMOS implementation. In 2012, H. Manem et al. [76] added another contribution to the field of neuromorphic computing using memristor based threshold gates by introducing a variation-tolerant training methodology to efficiently reconfigure memristive synapses in a Trainable Threshold Gate Array (TTGA) system. The TTGA consisted of arrays of trainable threshold gates [73] interleaved with switch blocks to enable the realization of the reconfigurable logic fabric as shown in Fig. 11. A single layer 4-input perceptron (trainable threshold gate) capable of implementing all linearly separable 1, 2, 3 and 4 input Boolean functions

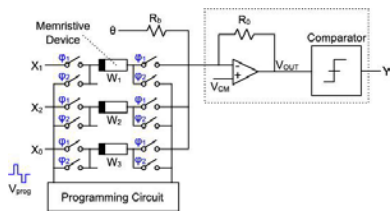

Fig. 10. Circuit implementation for programming memristive devices [74]

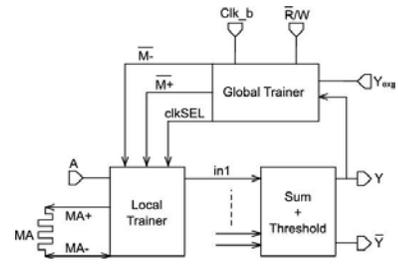

Fig. 11. Fully connected Perceptron with training circuitry [76]

[77] was considered to be the unit sized configurable logic block (CLB). The proposed TTGA system was designed and implemented from trainable perceptron based threshold gates [77] in Cadence Spectre with 45 nm Berkeley predictive technology models (PTM) for the CMOS circuitry. The proposed training methodology based on the stochastic gradient descent training technique was capable of efficiently reconfiguring the memristive synapses in a trainable threshold gate array (TTGA). The technique overcame many circuit level issues such as parasitics and device variations that configured memristive devices. The training and performance results for the TTGA and the 1T1M (1 Transistor and 1 Memristor) multilevel memristive memory [42], [78], [79] showed that the TTGA (minimum memristance values, i.e., OR pretrain) to be the most energy-delay and area effective solution. The methodology was observed as robust to the unpredictability of CMOS and nano circuits with decreasing technology sizes.

In 2013, L.Gao et al. [80] proposed that a programmable threshold logic gate can be implemented using a hybrid CMOS/memristor logic. The proposed linear threshold gate (LTG) shown in Fig. 12 was comprised of memristive devices in a universal gate which is much more powerful than similar fan-in single NAND or NOR gates. The memristive devices implemented a ratioed diode-resistor logic wherein several (N) memristive devices connected in parallel to a single pull-down resistor $R_L$, so that a dynamic range (i.e. the ratio $R_{ON}^H/R_{ON}^L$) dictated the number of different Boolean functions the LTG could implement. This configuration made LTG in-field configurable and potentially very compact.

The concept was experimentally verified by implementing a 4-input symmetric linear threshold gate with an integrated circuit CMOS flip-flop, silicon diodes, and Ag/a-Si/Pt memristive devices. For their effectiveness in high-throughput pipelined circuits, the CMOS flip-flop was preferred over a CMOS gate (and possibly an inverter) to restore the output voltage to a clear binary. The proposed implementation claimed to be more robust as compared to approaches suggested in [81] and [70], since it does not rely on changing the state of memristive devices during the logic operation. In addition, each memristive device in the suggested threshold logic in [70], [72], [82] was served by a CMOS-based current mirror circuit, which leads to considerable overhead. However, the memristive devices in the proposed hybrid architecture could be integrated into crossbar circuits which are patterned above the CMOS layer (e.g., similar to the proposed (3-D) CMOL concepts [43], [83], such that the threshold gate area would be determined mostly by the CMOS flip-flop. The method required only some minor additional circuitry for programming the memristive devices.

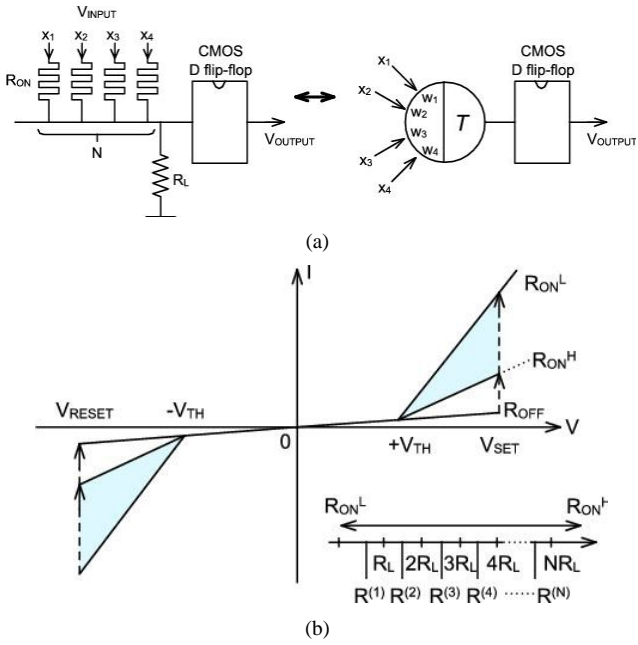

Fig. 12. **(a)** Main idea of an LTG implemented with memristors and CMOS D flip-flop. **(b)** IV characteristics of memristive devices (schematically represented). The shaded area on panel (b) shows the range of possible intermediate states utilized for the implementation of the threshold gate [80]

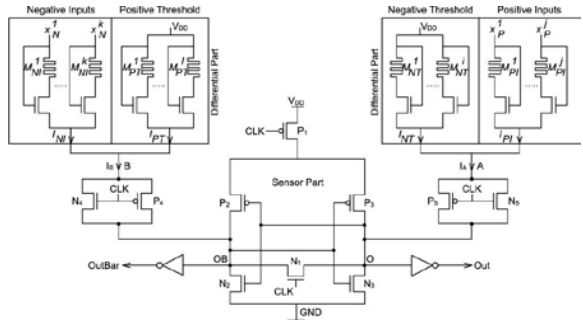

Fig. 13. Current-Mode Memristor based Threshold Logic (CMMTL) [84]

A new clocked design that combines memristors and CMOS transistors to implement current mode logic threshold logic gates was presented by C.B. Dara et al. [84] in 2013. The novel Current-Mode Memristor based Threshold Logic (CMMTL) shown in Fig. 13 consisted of two parts, a differential part and a sensor part, the differential being a series combination of a memristor and NMOS and having two further divisions comprising a negative threshold and positive inputs, and a positive threshold and negative inputs. Inputs are applied to the positive/negative parts of the differential part. When for a given input configuration the current $I_A$ through node A is greater than the current $I_B$ through node B, the voltage at output node $O$ rises faster than the voltage at output node $O_B$ resulting in high voltage at the node $O$ and low voltage at $O_B$. The inverse voltage allocation at $O_B$ and $O$ happens when $I_B$ is greater than $I_A$. The approach was found to outperform the combinational design [73] proposed earlier in terms of performance and energy consumption on three, four, and five input benchmark threshold functions. The percentage improvement were summarized as 77% for the delay, 50% for the energy and 88% for the Energy Delay Product (EDP). The delay and energy consumption using the proposed implementation using Berkeley Predictive Technology Models (PTM) for 45nm CMOS transistors were 0.44ns and 3.410, respectively. The proposed CMMTL is capable of implementing all possible weight configurations i.e., positive weighted gates, negative weighted gates, positive and negative weighted gates. The proposed method scales well over [73] as indicated by the increase in average EDP with the increase in number of inputs to the threshold function. Both the current mode logic of [84] and the combinatorial design of [73] used sense amplifiers to restore output voltage swing as opposed to CMOS D-flip-flops of [80].

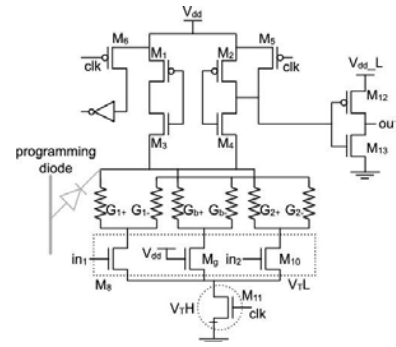

Fig. 14. Circuit for 2-fan in DRTL gate [85]

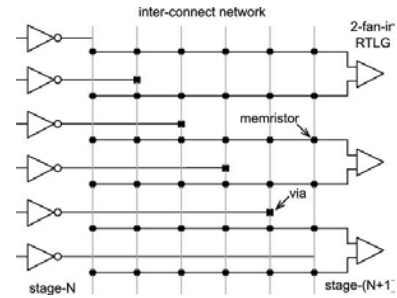

Fig. 15. Schematic for interconnect design for DRTL using resistive crossbar memory [85]

A Dynamic Resistive Threshold-Logic (DRTL) design based on non-volatile programmable resistive memory elements for reconfigurable computing was proposed by M. Sharad et al. [85] in 2013. In DRTL shown in Fig. 14, the resistive memory elements are used to implement the weights and the thresholds, while a compact dynamic CMOS latch is used for the comparison operation. The multiple stages in a DRTL design could be connected using energy-efficient low swing programmable interconnect networks based on resistive switches. The dynamic operation of the CMOS latches that minimizes static-power dissipation [43], [72], [74], [86] along with the memory-based compact logic and interconnect design

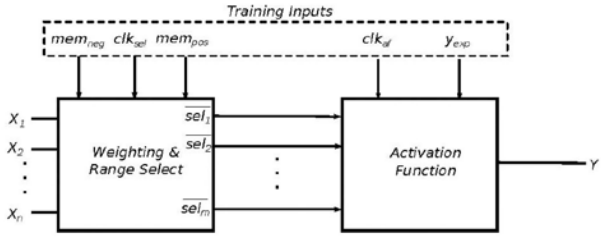

Fig. 16. Neural logic block design from [87]

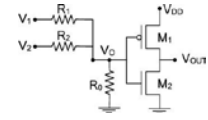

Fig. 17. Circuit diagram of the resistive divider boolean logic cell that consists of a two input resistive divider and a variable threshold CMOS inverter [88]

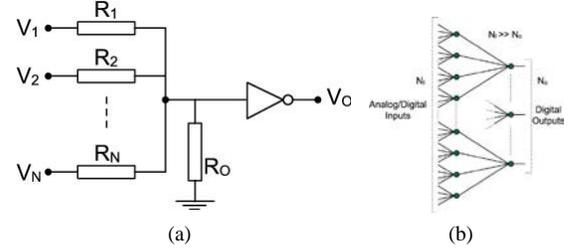

Fig. 18. Memory network cell and architecture **(a)** Cell structure **(b)** Network architecture (each node represents a cell) with no crossover wiring. Cells are arranged in hierarchical manner with N1 > N0 [88]

shown in Fig. 15 makes DRTL a dynamic, pipelined logic scheme with low power consumption and high performance. The performance analysis of DRTLG and interconnect design was evaluated by comparing the performance of DRTL with 4-input LUT based CMOS FPGA [33], for some ISCAS-85 benchmarks. The performance results show the possibility of 96% higher energy efficiency and more than two orders of magnitude lower energy-delay product for DRTL.

In 2013, Soltiz et al. [87] presented a robust and area efficient hardware implementation of a neural logic block (NLB) with an adaptive activation function, containing a Weighting and Range Select and an Activation Function, as shown in Fig. 16. The main motivation was to make threshold function adaptive so that nonlinearly separable functions such as XOR could be implemented within a single layer. Inspired by the principle of neuromodulators in the brain, the adaptive activation function comprised of $m$ points can model any function with $(m-1)$ boundaries. Weighting and Range Select component applies an adjustable weight to each input, calculates the weighted summation, and determines which of the $m$ ranges the summation falls in. The Activation Function associates a digital output with each range. The inputs are passed through memristors that are trained to a memristance M, ranging from $R_{on}$ to $R_{off}$. To change the functionality a different value of the memristance might be selected. The authors note that while the proposed design adds significant complexity when compared to a threshold activation function, the adaptive activation function provides benefits of fast training convergence times when compared to a neural logic block that only adjusts input weights, by training the shape of the activation function. On an optical character recognition (OCR) application the work shows a 90 percent improvement in the EDP over lookup table (LUT)-based implementations.

A universal Boolean logic cell based on an analog resistive divider and threshold logic circuit useful for mimicking brain like large variable logic functions in VLSI was proposed by A. P. James et al. [88] in 2014. The logic cell shown in Fig. 17 employed a CMOS - Resistance Threshold Logic co-design which successfully optimised the circuit design of conventional CMOS based large variable boolean logic problems. In the proposed resistance based threshold logic family, the resistive divider was implemented using memristors. The output of resistive divider was then converted into a binary value by a threshold operation implemented by CMOS inverter and/or Op-amp. For a two-input resistance divider circuit, if the threshold voltage of the inverter was set between 0V and 1/3V, the cell would work as NOR logic and if it was between 2/3V and 1/3V the cell would work as NAND logic. To operate the cell with a large number of inputs (>20) the threshold voltage of the inverter, for example in case of NOR, needed to be lowered to a very small range. To accommodate this effect, the authors introduced three inverters (Fig. 18) with three different $V_{DD}$s to form a universal gate structure to implement AND, NAND, OR, NOR, and NOT logic. For the cell to work as a NAND logic, the switches $S_2$ and $S_4$ were closed, and the output was taken from $V_{out}$ so that three inverters would be enabled. To implement AND logic, the switches $S_1$ and $S_3$ were closed, and the output taken from $\overline{V_{out}}$. If the switches $S_1$ and $S_4$ were closed, a NOR logic from $V_{out}$ was achieved: here only one inverter had to be enabled. If both $S_2$ and $S_3$ were closed, OR logic could be implemented. The proposed universal logic cell was based on the cognitive memory network [89], a resistive memory network that has no crossover wiring that overcame the hardware limitations to size and functional complexity associated with conventional analogue neural networks. The universal logic cell shown in Fig. 19 was employed to realize in an application to implement conventional digital logic gates. The simulation was performed in SPICE using feature size of 0.25m TSMC process BSIM models and HP memristor model for comparison with the CMOS implementation using a 16 bit adder and a 16 x 1 MUX. The analysis shows that the proposed cell offered advantages of smaller area and design simplicity in comparison with CMOS based logic circuits when the number

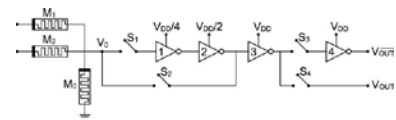

Fig. 19. Circuit diagram to implement NAND, NOR, AND, OR and NOT logic functions consisting of memristive resistance divider and CMOS inverters with three different power supply values [88]

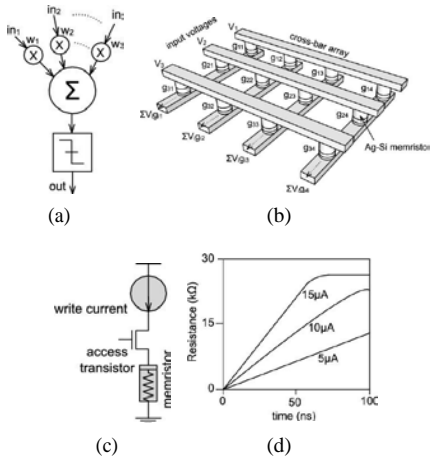

Fig. 20. **(a)** A Schematic representation of a threshold logic gate (TLG), **(b)** memristive cross-bar array, **(c)** A resistive memory cell with access transistors, **(d)** transient change in resistance for different magnitude of programming current [90]

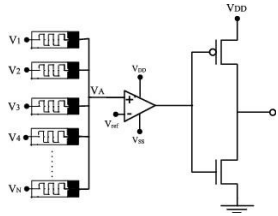

Fig. 21. Memristive threshold logic cell [93]

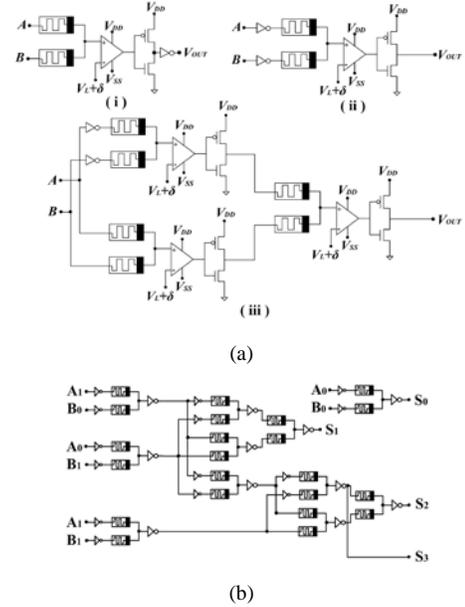

Fig. 22. **(a)** Circuit diagrams of the logic gates using proposed cell (i) OR gate (ii) AND gate (iii) XOR gate **(b)** 2 bit memristive threshold vedic multiplier [93]

of input variables became very high.

In 2014, Deliang Fan et al. [90] proposed a spintronic threshold device which can be combined with CMOS compatible Ag-Si memristors for designing ultra low energy Spin-Memristor Threshold Logic (SMTL). The SMTL gates shown in Fig. 20 employ memristive cross-bar array (MCA) to perform current-mode summation of binary inputs. The low-voltage fast-switching spintronic threshold devices (STD) shown in Fig. 28, based on magnetic domain wall, was found suitable for the design of energy efficient SMTL. The spin-torque switches based on magneto-metallic domain wall (DW) motion [91], [92] allows ultra-low voltage operation of memristive TLGs leading to low energy dissipation at the gate level. Field programmable SMTL gate arrays was found to operate successfully at a small terminal voltage of ~50mV, resulting in ultra-low power consumption in gates as well as programmable interconnect networks. The performance analysis done on common benchmarks show that the proposed hardware can achieve more than 100x improvement in energy and 1000x improvement in energy-delay product, as compared to the state of the art CMOS FPGA based TLG. Threshold logic computing using hybrid Memristive-CMOS cell architecture designed for Fast Fourier Transform and Vedic Multiplication have been proposed by James A.P. et al. in 2014 [93]. The proposed architecture involved a memristive threshold circuit configuration which consisted of the memristive averaging circuit in combination with operational amplifier and/or CMOS inverters as shown in Fig. 21 in application to realizing complex computing circuits. The developed threshold logic claims to outperform the previous memristive-CMOS logic cells by providing lower chip area, lower THD, and controllable leakage power; except for a higher power dissipation with respect to CMOS logic.

The proposed Memristive Threshold Logic (MTL) cell was designed specifically for implementing FFT and multiplication circuits as shown in Fig. 22 in modern microprocessors with the desired lower power dissipation and smaller on-chip area footprint using nanoscale fabrication techniques. Some additional desirable features of the proposed cell include the generalisation ability of the cell with a single cell structure with multiple functionality; and robustness to process variability in temperature, memristances and technology lengths indicating the fault tolerance ability of brain like logic circuits. The paper reports the successful application of the MTL cells in the examples of FFT and vedic multiplication computing circuits.

In 2014, Soudry et al. [94] proposed memristor-based grid to perform multiplication operation for learning backpropagation algorithms in multilayer neural networks (MNNs). Synapses comprising one memristor and two CMOS transistors were set in a grid formation each receiving two complimentary read/write pulses and an enable signal, and outputting on a current line. The column inputs fed the training data and row inputs the classification label. The two computational bottlenecks the authors addressed in gradient descent machine learning algorithms were matrix × vector and vector × vector multiplication. The matrix × vector product was implemented through the memristive grid by multiplication through Ohms law and analog summation of currents. The vector × vector product was done using time × voltage paradigm under the approximation that given a voltage pulse, the conductivity of a memristor would increment proportionally to the pulse duration multiplied by the pulse magnitude. The authors proposed that the area and power consumption were expected to be reduced by a factor of 13-50 in comparison with standard CMOS technology.

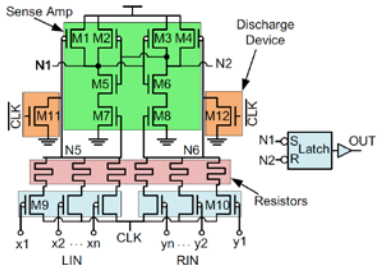

Fig. 23. Schematic of a DTG circuit [95]

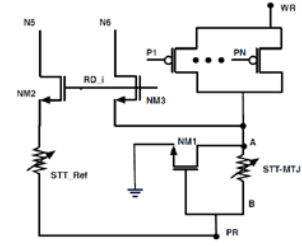

Fig. 24. STL-Cell [96]

Yang et al. in 2014 [95] used oxide based resistive RAM (RRAM) devices to implement threshold logic in order to compute logic functions with low power, robust circuits at low supply voltages (0.6 V). They used a differential threshold logic gate (DTG) circuit that consisted of five main components (Fig. 23): (1) a differential sense amplifier, which consists of two cross coupled NAND gates, (2) an SR latch, (3) two discharge devices, (4) left (LIN) and right (RIN) input networks, and (5) a network of resistors. The resistive network was implemented using oxide-based random access memory to behave as as CMOS compatible nano-scale resistor. The circuit outputs a logic 1 based on the inequality of number of active pFETs in the LIN and RIN networks. The benefits of robustness, area, and energy delay were demonstrated on a 16-bit full adder and a 128-bit comparator.

Earlier in 2012 Nukala et al. [96] from the same group had used an STT-MTJ (Spin Torque Transfer-Magnetic Tunnelling Junction) device with conventional MOSFETs to build threshold logic architecture. The resulting cell was used to program a large number of threshold logic functions, many of which would require a multilevel network of conventional CMOS logic gates. Based on an array architecture of these cells they demonstrated the advantages of non-volatility and zero standby power on a 16-bit carry look-ahead adder and compared with two conventional FPGA implementations. The array had 12x lower transistor count (compared to CLA-FPGA) and 10x reduction (compared to Ripple Carry-Ahead-FPGA) with comparable energy. The cell is shown in Fig. 24. In the *write* phase of the cell, WR is asserted, so that a certain amount of current (I), flows through the STT-MTJ device depending on the number of ON PMOS transistors. If $I \geq (Ic)$, the switching current, then the STT-MTJ device switches to the low resistance state otherwise it remains in high resistance state. When the WR pulse goes low, no current flows through the STT-MTJ device, and since the device is non-volatile, the state is maintained.

In 2014, A. K. Maan et al. [97] presented a programmable Memristive Threshold Logic (MTL) circuit design for real-time detection of moving objects. The proposed threshold logic shown in Fig. 25, which was targeted at high speed imaging

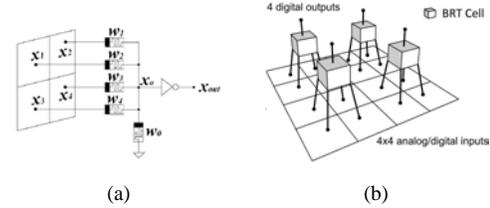

Fig. 25. **(a)** A modular Bilevel Programmable Resistance memristive Threshold logic (BPRT) cell that used 4-input pixel values **(b)** An architectural representation of the cell arrangement for a 4 × 4 pixels image; in this configuration, each cell uses 4 analog/digital inputs, has bi-valued weights, and implements image dimensionality reduction [97]

that could lead to near-continuous real-time object tracking and even surpass human object tracking ability, outperformed CMOS equivalent implementation in terms of area, leakage power, power dissipation and delay. The proposed resistive switching based threshold logic cell comprised of a voltage divider circuit and a CMOS inverter gate which encoded the pixels of a template image. The presented logic provided a framework to implement brain like logic in a memory and learning driven detection of multiple objects.

TABLE I
IMPLEMENTATIONS OF MTL - A COMPARISON

| Methodology | Area | Power (*mW*) |
|---|---|---|
| J. Rajendran et al. (2009) [70] | 77.52 $pm^2$ | 44.12 |
| J. Rajendran et al. (2010) [72] | 234.46 $pm^2$ | 32.56 |
| J. Rajendran et al. (2012) [73] | 193.77 $pm^2$ | 31.3 |
| T. Tran et al. (2012) [74] | 4.0 $nm^2$ | 30.8 |
| H. Manem et al. (2012) [76] | 193.87 $pm^2$ | 45.7 |
| L. Gao et al. (2013) [80] | 0.4992 $\mu m^2$ | 28.81 |
| C.B. Dara et al. (2013) [84] | 270.55 $pm^2$ | 118.22 |
| James A.P. et al. (2014) [88] | 155.52 $pm^2$ | 45.34 |
| James A.P. et al. (2015) [93] | 2.08 $nm^2$ | 30.72 |
| A. K. Maan et al. (2015) [97] | 38.91 $pm^2$ | 43.66 |

In Table I, the MTL implementations that were discussed have been compared on the basis of their area and static power dissipation. The analyses have been carried out in circuits by realizing a 2-input logic function using the proposed MTL techniques using simulations under same technology parameters (i.e. HP memristor model with 0.25u TSMC technology). According to Table II, [97] requires the lowest implementation area, which is 38.91 $pm^2$, and [80] has the lowest static power consumption, which is 28.81 mW. Among the 2-input cells implemented to process the image pixels [70], [97], the MTL design proposed by A. K. Maan et al. [97] in 2015 has the NR - Not Reported in the reference work. *Transistor count **Transistor count for C17-ISCAS-85 benchmark *** C432-ISCAS-85 benchmark; 'ISCAS-85 Benchmark C17-Before refresh " ISCAS-85 Benchmark C17-After refresh 'Writing to 1T1M memory circuit of physical array size 4 x 4; "Reading from 1T1M memory circuit of physical array size 4 x 4 ""Implementation of 3-input function "A+BC" using CMMTL lowest area and static power consumption. From the memory crossbar array implementations [85], [90] compared in Table I, the MTL design proposed by Deliang Fan et al. [90] in 2014 has the lowest energy-delay product.

TABLE II
COMPARISON OF THRESHOLD LOGIC IMPLEMENTATION USING MEMRISTORS

| Methodology | Area | Power | Leakage Power | Speed | Fan-in | Fan-out | Target application | Energy | Delay | Energy Delay Product (EDP) |
|---|---|---|---|---|---|---|---|---|---|---|
| J. Rajendran et al. (2009) [70] | NR | 0.57- 1.56$\mu$W | NR | NR | NR | NR | Image Classification | NR | 40 ps | NR |
| J. Rajendran et al. (2010) [72] | 28* | 42 $\mu$W | NR | NR | NR | NR | NR | NR | 6.1 ns | NR |
| J. Rajendran et al. (2012) [73] | 182** | NR | NR | NR | NR | NR | Boolean Logic Gates | 6.3fJ', 3.15fJ" | 2.99ns', 5.84ns" | NR |
| T. Tran et al. (2012) [74] | NR | NR | NR | >1 GHz | >10 | NR | Digital Logic Gates | NR | NR | NR |
| H. Manem et al. (2012) [76] | NR | NR | NR | NR | NR | NR | Memory | 3.82nJ | 189ps', 180ns" | 721.98zsJ', 687.6zsJ" |
| L.Gao et al. (2013) [80] | NR | NR | NR | NR | NR | NR | Digital Logic Gates | NR | NR | NR |
| C.B. Dara et al. (2013) [84] | NR | NR | NR | NR | NR | NR | Boolean Logic Gates | 3.41fJ"" | 0.42ns"" | 1.43sJ"" |
| M. Sharad et al. (2013) [85] | NR | NR | NR | 2 GHz | NR | NR | Programmable Logic Hardware, Crossbar memory | 480fJ*** | 0.5ns*** | 240ysJ |
| James A.P. et al. (2014) [88] | 70$pm^2$ | 9.2 $\mu$W | 0.014 nW | NR | 14.498M | NR | Boolean Logic gates, Arithmetic modules | NR | 0.45 $\mu$s | NR |
| Deliang Fan et al. (2014) [90] | NR | NR | NR | NR | 4 | NR | Programmable Logic Hardware | ≈300fJ*** | ≈2ns*** | ≈0.6zsJ |
| James A.P. et al. (2015) [93] | 4.55$mm^2$ | 3.00 $\mu$W | 14.30 pW | 1 GHz | NR | NR | FFT and Vedic Multiplication | 0.30pJ | NR | NR |
| A. K. Maan et al. (2015) [97] | 9.66$mm^2$ | 12.30 $\mu$W | 12.25 pW | 100 MHz | NR | NR | Object Detection | NR | NR | NR |

In Table II, a comparison between the various performance parameters of the threshold logic cells implemented with memristors that have been reported in the paper have been summarized. These example applications of threshold logic cells does not directly imply high degree of generality but indicate that the threshold logic circuits are scalable and can find its use in a wide range of tasks. The circuits are also arranged with respect of applications from the more general logic gate applications to very recent pattern recognition applications. Recent works indicate the power dissipation to be a concerning factor in the design of these circuits and require changes in the circuit configurations as well as memristor device design. The low area footprint for on-chip implementation has been the key attraction of these logic family of gates and circuits. The possibility that the threshold logic gates with memristors can lead to more-than-Moore's law outweighs the limitations of the power dissipation and lower speeds.

## IV. DISCUSSIONS AND OPEN PROBLEMS

The important properties of low leakage currents, ability of switching into memorised resistor levels, and smaller on-chip area make memristor an attractive element to mimic the principle of firing of neurons in silicon. Being a resistive device the biggest challenges in designing memristor based circuit configurations are with reduction in power dissipation in comparison with CMOS only circuits. The resistive path introduced by memristive devices within the designed threshold logic circuits often drives larger currents through the circuits resulting in higher power dissipation. Recently, the CMOS-memristor hybrid circuits have gained popularity as CMOS is a matured process technology, however, is gain-limited by higher power dissipation compared with CMOS-only counterparts. In addition, in MOS-gated memristive arrays the leakage currents become an important issue, that requires specialised compensated read-out techniques. Programmability of the memristive states in hybrid circuits become a challenging problem often requiring complex add-on circuits. If not designed carefully the area of programming circuits can outweigh the benefits of lower area offered by the memristors.

The progress in the VLSI circuits development depends also on the availability of accurate models for memristors, that can be incorporated into widely accepted simulation tools. Although there are several simulation models available in SPICE and VerilogA, they still do not cover the wide range of memristive devices available for use today. In addition, the physical level design of memristors and memristor-CMOS hybrid circuits is challenged by the lack of process standards and requires further adoption of these devices in semiconductor industry. In this sense, memristor devices and circuits are quite early in the stages of commercial implementation. We note that while there are aforementioned issues, the ability of memristive threshold logic cells to be general and programmable makes it an interesting alternative to CMOS logic family of gates. In addition, the ability of the threshold logic cells to be programmed to behave like a pattern recognizer mimics the principle of neurons that can provide multiple functionalities with the same cell structure.

The hardware programmability opens up a wide range of applications for circuit designers and artificial intelligence researchers. Some of these are: (1) development of memories that can learn, store and forget like human memory, (2) development of human like self-learning classifiers in silicon, (3) building higher level intelligent modules on programmable and self-learning chips, (4) building array processing circuits and re-inventing analog sensory processing such as in computer vision, (5) developing CAD tools and systems for memristive hybrid circuits, and (6) development very large scale simulation and implementation of such threshold logic systems.

Memristive behaviour is observed in devices dating back to electric arc experiments performed by Sir Humphry Davy in 1808. Although lost in the history, the resurgence of memristive device as a possible frontrunner in implementation of neural network is made possible mainly due to the advances in nanoscale technologies in last decade. In addition, the limitations imposed by scaling issues of MOSFET devices, and the need to create large scale neural networks for imitating silicon brain continue to inspire research in this field.

## V. CONCLUSION

This survey discussed the impact of memristive threshold logic by sketching the number of applications that have sprung up utilizing the inherent nature of several devices that show memristive behaviour. The paper reviewed the different circuit implementations of threshold logic that employed the resistive switching nature of memristors potentially leading to highly efficient, high density neurocomputing devices in future. Even with several new advances, this field of research is relatively very new, and offers several challenges and opportunities for neural networks, learning systems, circuits and systems communities to work together. The practical applications of memristive threshold logic design extend to real-time processing and recognition of natural signals, its differentiation and efficient architectures for silicon memories. While the possibilities are many there exist several challenges as well, including the need for better energy and power dissipation ratings, newer circuits for efficient programming of memristor arrays, accurate simulation models for memristors, and long-term development of foundry support.

ACKNOWLEDGEMENT

We would like to thank all the reviewers for the constructive feedback and comments, that has helped us to improve the overall quality of the paper.

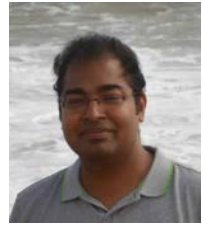

**Alex Pappachen James** works in the area of brain inspired circuits, algorithms and systems, and has a PhD from Queensland Micro and Nanotechnology Centre, Griffith University. He is currently an Associate Professor in Electronics Engineering at Nazarbayev University, and leads the Bioinspired microelectronic systems lab. His group is actively engaged in the development of memristive circuits and systems that can be integrated to realistic imaging and pattern recognition systems. He is a Senior Member of IEEE.
http://www.biomicrosystems.info/alex

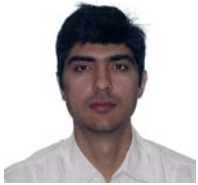

**Akshay Kumar Maan** completed his Ph.D. in Memristor-based threshold logic cognitive circuits at Griffith School of Engineering, Griffith University, Brisbane, Australia in 2015. The work is inspired by the principle of firing of neurons in the primate brain, and the ability of neuronal structures to perform different cognitive tasks, explores programmable (mem)resistance-based threshold logic memory networks that are shown to perform basic logic functions to highly involved cognitive tasks such as face recognition and fast moving object detection, in a parallel, scalable hardware architecture. His main areas of interest are brain inspired circuits, algorithms and systems, to enable more-than-moor era of low power, high density VLSI circuits for cognitive applications such as pattern recognition, face recognition and speech recognition.

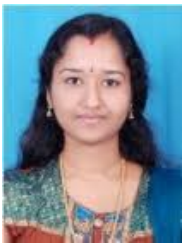

**Deepthi Anirudhan Jayadevi** received the B.Tech. degree in Electronics and Instrumentation Engineering from Cochin University of Science and Technology, India in 2008. She received the M.Tech. degree in VLSI and Embedded Systems from ER&DCI Institute of Technology (ER&DCI-IT), an educational institution under the direct control of Centre for Development of Advanced Studies (CDAC), India in 2010. She was a research staff member at Enview R&D labs, and is a member of the IEEE and has more than five years of experience in teaching and research in the field of Electronics. Her current research interests include VLSI circuits and artificial intelligence.